%% file: Porter.tex
\documentclass[graybox]{svmult}

\usepackage{mathptmx}       
\usepackage{helvet}         
\usepackage{courier}        
\usepackage{type1cm}        
\usepackage{makeidx}         
\usepackage{graphicx}        
\usepackage{multicol}        
\usepackage[bottom]{footmisc}

\makeindex             

\begin{document}

\title*{The Challenges in Gravitational Wave Astronomy for Space-Based Detectors.}
\author{Edward K. Porter}
\institute{Fran\c{c}ois Arago Center, APC, Universit\'e Paris Diderot, CNRS/IN2P3,
CEA/Ifru, Obs. de Paris, Sorbonne Paris Cit\'e,
                10, rue Alice Domon et L\'eonie Duquet, 75205 Paris Cedex 13, France, \email{porter@apc.univ-paris7.fr}}
%
%
\maketitle

\abstract{The GW universe contains a wealth of sources which, with the proper treatment, will open up the universe as never before.  By
observing massive black hole binaries to high redshifts, we should begin to explore the formation process of seed black holes
and track galactic evolution to the present day.  Observations of extreme mass ratio inspirals will allow us to explore galactic 
centers in the local universe, as well as providing tests of General Relativity and constraining the value of HubbleÕs constant.  The detection of
compact binaries in our own galaxy may allow us to model stellar evolution in the Milky Way.  Finally, the detection of cosmic
(super)strings and a stochastic background would help us constrain cosmological models.  However, all of this depends on our 
ability to not only resolve sources and carry out parameter estimation, but also on our ability to define an optimal data analysis
strategy.  In this presentation I will examine the challenges that lie ahead in GW astronomy for the ESA L3 Cosmic Vision mission,
eLISA.}

\section{Introduction}
\label{sec:1}
Gravitational waves (GWs) are a prediction of Einstein's theory of General Relativity and can be thought of as ripples of spacetime itself, which are caused by the high acceleration of massive objects.  This can be from the motion of a binary of compact objects, a supernova or the creation of the Universe itself.  GWs share many common points with electromagnetic (EM) waves.  They have two polarisations, which instead of a 90 degree rotation as is seen with photons,  are rotated to each other by 45 degrees.  Rather than being a dipolar radiation, GWs are a quadrupolar radiation.  They obey the inverse square law and have a propogation speed of that of light.  The difference is that whereas photons are oscillations of the EM field, GWs are oscillations of the fabric of spacetime itself.  The other main difference is that photons are scattered and diffracted as they make their way towards the Earth. As gravity is the weakest of the four fundamental forces, GWs interact very weakly with matter, meaning there is almost no scattering or diffraction.The weakness of the GWs makes their detection a very difficult task.  However, due to their low environmental interaction, they travel from source to detector with virtually no information loss.  This allows us to investigate the sources of GWs with unprecedented accuracy

In the next two decades, GWs will open a new window on the universe, providing the field of astrophysics with an unprecedented richness of sources.  The ground-based interferometers Advanced Virgo and Advanced LIGO~\cite{LIGO,Virgo} will come on-line in 2015 and operating at frequencies greater than 
one Hertz, 
will be sensitive to stellar mass compact objects in the $(1-10) M_{\odot}$ mass range.  In the nanohertz regime, researchers are investigating the detection of GWs from supermassive black hole binaries (SMBHBs) with masses of between $(10^8-10^{10}) M_{\odot}$ via pulsar timing measurements~\cite{pta}.  It is highly possible that both methods
 will make the first direct GW detection of astrophyisical sources before the end of the current decade.  In addition, ESA has recently chosen the theme of the ``Gravitational Wave Universe" for the ESA Cosmic Vision L3 mission~\cite{eLISA} which will explore the millihertz regime.  This frequency band, which is inaccessible to
  the other experiments, will explore SMBHBs with masses of $(10^3-10^7) M_{\odot}$.

An efficient way to detect GWs is by using interferometry.  The passage of a GW through an interferometer of length $L$ changes the distance between space-craft by an 
amount $\Delta L$.  The strain amplitude induced by the passage of the GW is given by $h\sim\Delta L/L$.  In general, nature provides a natural strain on the order
of $h\sim10^{-21}$.  As the GW changes the distance between space-craft by approximately 20 picometers, we can see that a space-craft
separation of $L\sim 10^{-12}/10^{-21}\sim10^6$ kms is needed to detect the GW.

The main detection technique for GWs is based on optimal or Weiner matched filtering.  The matched filter is the optimal linear filter in the presence of stochastic noise.  In most cases, we assume that the GW source is buried in instrumental noise and is
not visible in the time domain.  Matched filtering works perfectly in this case as it allows us to detect the source by transforming to the Fourier domain and using
a theoretical waveform model, or template, to correlate with the data.  By adjusting the parameters of the template, we can optimise the signal to noise ratio between
the template and a possible signal in the data.  The downside of matched filtering is that it requires a theoretical model of the possible signal that we are trying
to detect.  Furthermore, it is mostly dependent on the phase matching between a template and signal, so the technique requires very accurate phase modeling of the
waveforms.

In this lecture, we will investigate the main GW sources of interest for a space-based interferometer, listing the current state of the art and outstanding problems
in both astrophysics and algorithmic development.

\section{Galactic Binaries}
The eLISA data stream will be source dominated with approximately 60 million GW sources.  Approximately 
$99\%$ of these sources will be Milky Way binaries composed of white dwarf, neutron star and stellar mass black holes.
While made up of compact objects which are described by General Relativity, these binaries can be thought of as being Newtonian in terms of their orbits (i.e. the orbital velocity of the binary components is very small compared to the speed of light).  Due
to the wide separation of the system components, these binaries are essentially monochromatic with simple waveform
polarization models.

While  most of these galactic binaries will be monochromatic, we do expect a certain number of interacting binaries where we 
have mass transfer due to Roche lobe overflow.  In these cases, we expect to be able to measure the first frequency derivative for a certain number
of sources, and for a smaller number of sources, possibly even the second frequency derivative.  A measurement of the frequency derivatives would allow
us to break certain parameter correlations and improve parameter estimation.

For the LISA mission it was estimated that we should be able to individually resolve approximately $25\times10^3$ sources~\cite{Rubbo}.  For
the eLISA mission this number falls to around $3500$ sources with one year of observation~\cite{eLISA}, with an additional $10^3$ sources
per year~\cite{Tyson}.  This is due to the fact that the frequency resolution $\Delta f$ goes as the inverse of the observation time $T_{obs}$.  Therefore as
$T_{obs}$ gets larger, the width of each frequency bin $\Delta f$ gets smaller allowing us to resolve more sources.  While the total number of resolved binaries decreases
with the eLISA mission, one of the benefits is the lack of a confusion problem.  This arises from the fact that many thousands of galactic binaries occupy
the same Fourier bin.  Only the brightest sources in each bin can be resolved, leaving behind many unresolved sources per frequency bin.  
With LISA, even after the subtraction of $25\times10^3$, the background of unresolved sources remained above the noise curve.  This excess galactic noise could then effect the parameter estimation of other 
sources in the data.  For eLISA, subtracting the 3500 galactic binaries reduces the galaxy to the instrumental noise level, eliminating the galactic
confusion problem. 

\subsection{Verification Binaries}
These are compact binaries in the Milky Way for which we already have EM information.  These are guaranteed sources for an ESA L3 mission.  Due to the fact that
we already have information on these binaries, they should also be the first sources we detect.  As well as the fact that they are guaranteed sources, these binaries will also 
allow us to recombine the data sets in a coherent manner in case of a disturbance or disruption of the data, due to the inherent knowledge of the frequency of the 
binary.

At present there are less than 10 known verification binaries.  However, in 2011 a new white dwarf binary was discovered with an orbital period of 39 minutes
~\cite{Kilic}.  This binary, labelled J0106-1000, has a GW frequency right in the middle of the eLISA band.  It is also hoped that GAIA~\cite{GAIA} will discover a number of 
additional verification binaries.  While GAIA will not have the resolution to provide full parameter estimation, it is possible that follow-up observations with other
EM telescopes will provide many tens, if not hundreds, of new verification binaries by time a mission like eLISA launches.

\subsection{Current status of existing algorithms}
Only a small number of algorithms exist worldwide to search for galactic binaries~\cite{Crowder, Blaut, Littenberg}.  These algorithms were based on Markov Chain Monte Carlo or stochastic template bank algorithms.  The first of these algorithms uses a stochastic search process to improve the parameter estimation of the source by moving to points of higher likelihood in the parameter space.  The algorithm can be accelerated by using techniques such as simulated annealing (i.e. the 
likelihood surface is heated to remove all but the highest peaks, and is then slowly cooled back to a normal state) and by the use of source specific proposals when
making the jump to the next point in the parameter space.  The second algorithm is more of a brute force method where the parameter space is covered in a 
grid of templates, and each template is correlated with the data set.  Points of high correlation are then investigated further.  
However, since 2010 there has not been much
further development in galactic binary algorithms.  While the algorithms were applied to Mock Data Challenges~\cite{mldc} with varying levels of success, they were never applied to the reformulated 
eLISA mission.  However, as the number of resolvable binaries has decreased, if anything, the existing algorithms should be more performant.

\subsection{Outstanding problems}
As with all the sources, it is possible to split the outstanding problems into two branches : astrophysical and algorithmic.  
\begin{enumerate}
\item We still do not have a solid grasp
on stellar evolution in the Milky Way.  In a standard evolutionary picture, we know that the compact remnant of a supernova is imparted with a spatial kick
velocity.  If big enough, this kick velocity could disrupt the binary.  For a compact binary to form, the system would need to survive the kick velocities from two
supernovae.    Furthermore, as we do not fully understand the initial mass distribution of stars in the Milky Way, it is very difficult to estimate the distribution of 
binaires in terms of those composed of white dwarves, neutron stars and stellar mass black holes.  We would also expect there to be a certain fraction of mixed
systems.

\item On the algorithmic side, the development of search and resolution algorithms needs attention.  While it has been shown that there is no confusion between a 
single galactic binary and an other source type, such as a massive black binary, we have seen that the galaxy of white dwarf binaries can fool an algorithm
into believing it has found a massive black hole binary (this is the so-called white dwarf transform).  However, this can be circumvented using a Bayesian model
analysis.  The big question that needs to be answered is can we search for galactic binaries without removing the other brighter sources first?  This should be 
a priority line of study in the coming years.  

\item Finally, the current algorithms model galactic binary sources as circular binaries.  However, we know that the famous
Hulse-Taylor binary B1913+16 has an eccentricity of $e\sim0.6$~\cite{Hulse}.  While it is well known that GWs are very efficient in circularising a binary, it may be
that a large fraction of the galactic binary sources will have a non-negligible eccentricity.
\end{enumerate}

\section{Supermassive Black Hole Binaries}
A principal GW source for a space-based mission is the merger of supermassive black hole binaries (SMBHBs).  The luminosity of a SMBH merger in GWs is around $10^{26} L_{\odot}$, making them the most violent and brightest events in the Universe (In comparison, a supernova releases $\sim10^{14} L_{\odot}$ in photons 
and generally outshines the host galaxy).  As the waves travel towards the detector almost untouched, it allows us to see the source almost as it was and estimate the parameters of the binary system with incredible accuracy.  It is estimated that we will be able to measure the component masses to less than 1\% error (see for example Fig 3 of Ref~\cite{eLISA} - a measurement of astrophysical masses on a level never seen before), the luminosity distance with an error between 1-50\% and estimate the time of merger to within a few minutes. The expected event rate for these sources is $\sim$100/yr~\cite{Sesana07,Sesana11}. Detection of these sources would allow us to map the black hole mass function and the spin history of the binary as a function of cosmic time, observe the mechanisms for the formation of massive black holes during the dark ages of the Universe and test alternative theories of gravity during the inspiral, merger and ringdown of the binary. 

There is almost unquestionable evidence that SMBHs lie at the center of each galaxy.  There is also EM evidence for the existence of massive black hole binaries, although we should mention that resolution of these sources is incredibly difficult, so a note of caution should be aired.  What is certain is that we know that galaxies collide and that the SMBHs that we see today did not form as they are observed.  The working hypothesis for black hole growth is that the seeds of SMBHs formed at high redshifts $(10 \leq z \leq 17)$,  and via subsequent galactic mergers, grew to the size they are observed at today.  These merger tree models(see~\cite{Volonteri10} for example) seem to give a reasonable explanation of the Universe we observe today.  However, one of the main questions in observational cosmology and astrophysics is how did the first massive black holes form.  

At present there are two main hypotheses~\cite{Volonteri10,Begelman06,Devecchi12,Madau01,Mayer06,Tegmark97} : the first is that the first black hole seeds came from the remnants of Population III stars.  These were very massive, low metal stars with very short lifetimes ($\sim$ million years).  The resulting supernova would produce black holes with masses of between $(10-100) M_{\odot}$.  The second hypothesis is that some protogalaxies underwent direct gravitational collapse forming black holes with masses of between $(10^3-10^4) M_{\odot}$.  However, no observational evidence exists for either scenario.

\subsection{Algorithmic Development}
As they are a priority source, the area of development of search algorithms for SMBHBs has probably seen the highest activity in the last decade.  A number of algorithms of different types have been developed to search for SMBHBs.   Each algorithm was successful at a certain level, but through the MLDCs it was realised that many of the algorithms had fundamental limitations.  The developed algorithms followed a number of different strategies.  Some groups tried to apply the template grid method that was developed in the ground-based community.    These groups either tried a multi-stage algorithm~\cite{Brown} or a stochastic template bank~\cite{Babak}.  Both of these algorithms detected the source and had varying success in estimating the parameters of the source.  It is not clear how well these algorithms will work in the future when applied to realistic sources in a full sized parameter space.  In general, the number of templates for a template bank grows
geometrically with the dimensionality of the parameter space.  If we assume that one day we will search for spinning, eccentric binaries, then the bank needs to cover
a 17 dimensional parameter space.

Other groups applied a time-frequency algorithm (see ~\cite{mldc} for example) where short duration Fourier transforms of the data are used to identify 
coherent tracks in the time-frequency plane.   As the phase information of the wave is lost, it makes it very difficult to carry out parameter estimation using this
technique and imposes a fundamental limit of the usefulness of the algorithm.

A third group of algorithms were developed with the idea of conducting the matched filtering in a more sophisticated manner.  These were based on adapted
Metropolis-Hasting Markov Chains~\cite{cp1, cp2, cp3, cp4}, Nested Sampling~\cite{feroz}, Genetic Algorithms~\cite{petiteau} or Evolutionary Algorithms~\cite{hea}.  These algorithms were very
successful in the detection of SMBHBs, and in most cases estimating the parameters of the source.   The main difference between the algorithms was that the
Metropolis-Hastings algorithm searches for single sources in series, while the Nested Sampling, Evolutionary and Genetic Algorithms search for multiple sources at the same
time.

\subsection{Outstanding Problems and Questions}
While a lot of progress has been made in astrophysical modeling, GW theory and GW astrononmy in the last decade, a number of outstanding questions still remain.
\begin{enumerate}
\item There still is no solution to the question of how the sources should be searched for.  It could be that is possible to search for the SMBHBs without having to worry about the other source types, in which case any of the above style of algorithms will work.  However, if this is not true, and we have to search for sources in terms of their signal to noise ratio, then it maybe makes more sense to develop algorithms that only search for single sources.
\item The current studies have yet to come close to an astrophysically realistic data set in terms of source numbers.  It is estimated that there could be anything up to
100 SMBHBs per year in the data stream.  For the cases already studied, a solution exists for the restricted case of a small number of Schwarzschild SMBHBs.  The community has begun to investigate the case of Kerr SMBHBs, but a solution does not yet exist.
\item  In the previous studies, the priors on the parameter values have been quite narrow.  It is unclear how the current algorithms would perform on open
priors.
\item While we would like to detect every merger possible, the question arises as to do we need to do this in real time?  If we believe that there may be EM counterparts to the merger, then it is clear that we would like to inform the wider astrophysical community.  However, there may be a luminosity cutoff which means that not every
merger will have a detectable EM signature.  These systems could then be treated at a more relaxed pace.
\item Do we expect to see EM counterparts for SMBHBs?  It could be that even if these processes are super-Eddington in nature, the source is simply too distant to observe
the counterpart.
\item How do gas and accretion disks effect the evolution of spin in SMBHBs?  We know that in gas rich environments, the spins of the black holes align with the
total angular momentum of the system.  In this case, we could use a simplified template to search for SMBHBs.  
\item What are the formation processes of massive black holes at high redshift?  It is possible that nature will present us with a mixture of high and low mass seeds.
However, none of the current astrophysical models explain the presence of hypermassive quasars at high redshift.  Because of these uncertainties in the models, the 
actual expected event rate is very difficult to calculate.
\item Do we have realistic templates?  I will discuss this below.
\end{enumerate}

\subsection{Comparable Mass Waveforms}
As stated earlier, the main detection technique in the search for GWs is matched filtering.  When developing a template, we need to have a strong theoretical 
knowledge of the waveform model if we are to carry out an accurate estimation of parameters.  Up to now, pretty much all algorithmic development and parameter
estimation studies were conducted using circular Schwarzshild templates.

In reality we know that these waveform models will be good enough to announce a detection, but will not be good enough for parameter estimation.  It was demonstrated that the effects of spin and spin precession break correlations between parameters and improve parameter estimation~\cite{Lang}.  There is a further
breaking of parameter correlations when one includes higher harmonic corrections to the waveforms.  It was shown that these corrections provide a massive improvement in both the luminosity distance and the sky position of the source~\cite{hhc1, hhc2, hhc3}.  While we know that GWs are very effecient at circularising
binaries, it was shown that three body interactions in stellar clusters and galactic centers can push the eccentricity of the system so high that it enters the detection
band with a non-negligible value.  In this case, it is extremely difficult to properly estimate the parameters of the source using circular templates~\cite{ecc1, ecc2}. 

In many cases, the merger and ringdown of the waveform have also been neglected.  However, with advances in Numerical Relativity, not only are longer
waveforms being produced, but analytic waveforms that are calibrated to the numerical results are also being produced~\cite{eob1, eob2}.  

Ultimately, when a space-based GW observatory launches, we would like to have templates that include eccentricity, spins, higher harmonic corrections, merger
and ringdown.  This presents a major theoretical challenge for the next decade.

\section{Extreme Mass Ratio Inspirals}
Another major source of GWs are the Extreme Mass Ratio Inspirals or EMRIs.  This is a system composed of a stellar mass compact object obiting a central SMBH.  This object could be anything from an ordinary star to a stellar mass black hole.  It is assumed that before reaching the true strong field regime close to the
central black hole, a stellar mass star would be tidally disrupted, and compact objects such as white dwarves and neutron stars may undergo a thermonuclear
detonation from tidal heating due to the eccentric orbits.  It is therefore believed that the main observable systems will be composed of a stellar mass black hole
orbiting a SMBH.

EMRIs are considered to be a very exciting source.  It is expected that we will see approximately 50 per year out to a redshift of $z\sim0.7$~\cite{eLISA, Gair}.  As the stellar mass object spends the last 1-2 years of its life in the strong field regime of the central black hole (1-3 Schwarzschild radii), these objects are an incredible opportunity to conduct spacetime cartography.  In this final 1-2 year period, we expect the smaller body to trace out between $10^4-10^6$ orbits in the vicinity of the central black hole, allowing us to confirm that it is a Kerr black hole as predicted by General Relativity, and not something more exotic like a Boson star.

\subsection{Algorithmic Development}
The EMRIs in general have received the least amount of attention in terms of development.  Two main algorithms exist.  These algorithms were adapted from
the SMBHB search codes and are essentially based on a Markov Chain Monte Carlo type algorithm.  During a mock data challenge, four out of a possible five
EMRIs were detected with good parameter estimation.  However, once again, this is nowhere near the expected astrophysical rate of 50/yr.  Furthermore, the
searches took place with very narrow priors.

The main problem with the development of algorithms for EMRIs is the run-time with current computational power.  The other sources are easier to develop for, as
an algorithm can be set running and the user knows within a couple of hours if things are going well or not.  This is a luxury we do not have with the EMRIs.  It usually
takes a few days before one can say whether or not there is a problem.  This slows down the algorithmic development dramatically.

\subsection{Outstanding issues}
While the list is shorter, there is no doubt that the following problems present an immense technical difficulty.
\begin{enumerate}
\item The question of whether or not we have realistic templates arises once more.  All of the studies and challenges conducted for EMRIs have used a 
phenomenological waveform that contains the necessary complexity (i.e. the correct number of parameters).  However, EMRIs are very difficult to model.
They have three fundamental frequencies and have very complex orbits.  While progress has been made on including self-force calculations (see for example~\cite{barack}),
it is not clear how close we are to having realistic EMRI templates.

\item As the current algorithms are based on phenomenological waveforms, their success was based on the analytic knowledge of the separation between
harmonics and sub-harmonics.  This is something that will not be known with realistic waveforms and might affect the performance of our algorithms.

\item Is there an EMRI confusion problem?  For those systems with little frequeny evolution in each harmonic, a real data set may contain multiple systems
that contain a harmonic in the same Fourier bin.  This could cause problems in identifying the source.

\item Will imperfect removal of a galactic binary affect the detection of an EMRI?  Again for those sources with very little frequency evolution, an EMRI can
theoretically be modeled by a collection of monochromatic sources.  It is not inconceivable that an EMRI harmonic could be mistaken for a galactic binary
and accidentally removed.

\item  What are the EMRI formation channels?  Do we require purely centrophilic orbits or can we expect the birth of compact objects in an accretion disk around the central black hole?  One can also approach this question from the other direction.  If we detect EMRIs, what information can we provide on the formation channel?

\item Can we deal with wide priors and astrophysically realistic event rates?  In theory, there is no reason to conduct an EMRI analysis in real time as we do not expect
to have an EM counterpart.  Further research is necessary to estimate how many EMRI detections are needed to answer any scientific questions that we deem important.

\end{enumerate}

\section{Cosmological Sources}
With the recent BICEP 2 result~\cite{bicep}, there is a large current interest in GWs from cosmological sources.  While the GWs discovered by BICEP 2 are not in
the same frequency band as eLISA, there are many possible sources that will be.  Two such sources are first order phase transitions and networks of cosmic
(super)strings.  In a previous MLDC~\cite{mldc} the community investigated the possibility of detecting bursts of GWs from cusps of cosmic (super)strings.  While
each source was detected, the parameter estimation, and in particular the localisation of the source on the sky, was not very good.  

This was mostly due to how a space-based observatory attains a sky position for the source.  For long lived sources, such as SMBHBs, the motion of the detector
around the sun induces a Doppler motion effect into the waveform.  Because the detector sees the source differently at different points in the orbit, it allows us to 
fix the source in the sky.  However, for short transient events like GW bursts from cusps, the detector has hardly moved during the lifetime of the source.  This means
that to get the sky position, we need to use triangulation between the space-craft.  This is much less accurate than using the Doppler motion and results in an
inferior estimation of the distance.

\subsection{Outstanding issues}
\begin{enumerate}
\item Very little development has gone into algorithms to detect a stochastic cosmological background.  It is unclear if one needs a three-arm detector to detect
such a background, or if it is possible with a two-arm detector.

\item There is a prediction that a network of cosmic (super)strings will produce a background that would be the dominant source of GWs for a mission such as
eLISA~\cite{Pierre}.  In this case the stochastic background would become the foreground source.  The question then arises if it is still possible to detect the astrophysical sources that would lie below such a foreground?

\item It is hoped that cosmological theories will be more constrained in the next decade.  This theoretical work would allow a better prediction of the strenght
of a possible stochastic background, as well as the possible event rates for transient bursts of cosmological origin.
\end{enumerate}
\section{Conclusion}
This is an exciting time in the field of GW astronomy.  The recent BICEP 2 announcement of the discovery of GWs from the early universe has brought the subject to the forefront of both the research and public eye.  With both
Advanced LIGO and Virgo due to come on-line in the next couple of years and the advance in pulsar timing arrays, there is a real possibility of the first detection
of GWs from astrophysical origins before the end of the current decade.  The recent decision by ESA to chose the Gravitational Wave Universe as the theme of the 
Cosmic Vision L3 mission would allow the community to cover the GW spectrum from $10^{-9}$ Hz up to $\sim$ kHz.

The last decade has seen a massive improvement in the modelling of astrophysical sources of GWs.  From advanced computer modeling of accrection disks and gas interactions, to merger tree models, to three body interactions, our understanding of the astrophysical universe has never been better.  The discovery of new binary systems with EM parameter estimation has now begun to increase the number of expected verification sources in our own galaxy.  In the next decade we hope to see
the discovery of many more of these systems and possibly even the concrete detection of SMBHBs with sub-milliparsec separations.

Since 2004, the field has also witnessed giant leaps in the development of search algorithms and statistical techniques for space-based GW sources.  For some sources,
the limit to our development is simply the computational power available to us today.  The community has developed sophisticated techniques that now give us 
confidence in our ability to carry out GW astronomy for objects such as SMBHBs, EMRIs and galactic binaries.

However, a lot of work still lies ahead.  A constant improvement in the efficiency and accuracy of our algorithms is needed.  More sophisticated templates incorporating
the maximum of physical effects need to be developed for all sources.  Better constraints on astrophysical event rates and cosmological models will be needed to 
guide the development of the field of GW astronomy.

While the next decade will be challenging,  it will also be very rewarding.  A new field of astronomy is opening up before us, and promises to provide us with a 
view of the universe never seen before.

\begin{acknowledgement}
The author would like to thank the organisers of the Sant Cugat Forum on Astrophysics for their invitation to give this lecture.
\end{acknowledgement}
%

\input{Porterreference}

\end{document}

%% file: Porterreference.tex
%
%
%
\def\prd{{Phys. Rev. D }}        
\def\cqg{{Class. Quantum Grav. }} 
\def\ekp{Porter,E.K. }
\def\njc{Cornish,N.J. }
\def\jrg{Gair,J.R. }